\documentclass{aa}
\usepackage{natbib}
\usepackage{epsfig}
\usepackage{csquotes}
\usepackage{tikz}
\usetikzlibrary{decorations.pathmorphing}
\usetikzlibrary{decorations.pathmorphing,calc}
\def\ut#1{\mathop{\vtop{\ialign{##\crcr
     $\hfil\displaystyle{#1}\hfil$\crcr\noalign
     {\kern1pt\nointerlineskip}\hbox{$\hfil\sim\hfil$}\crcr
     \noalign{\kern1pt}}}}}

\def\undersymbol#1#2{\mathop{\vtop{\ialign{##\crcr
     $\hfil\displaystyle{#2}\hfil$\crcr\noalign
     {\kern1pt\nointerlineskip}\hbox{$\hfil#1\hfil$}\crcr
     \noalign{\kern1pt}}}}}

\usepackage{graphicx}
\begin{document}

\title{Cosmic voids and the kinetic analysis. V. Hubble tension, the cosmological constant, and aperiodic filaments}
       \author{V.G.Gurzadyan\inst{1,2}, N.N.Fimin\inst{3}, V.M.Chechetkin\inst{3} }

              \institute{Center for Cosmology and Astrophysics, Alikhanian National
Laboratory, Alikhanian Brothers str.2, 0036, and Yerevan State University, Manukian str.1, 0025 Yerevan, Armenia \and
SIA, Sapienza Universita di Roma, Via Salaria 851/881, 00191 Rome, Italy \and Keldysh Institute of Applied Mathematics of RAS, Miusskaya Sq. 4, 125047 Moscow, Russia}

   \offprints{V.G. Gurzadyan, \email{gurzadyan@yerphi.am}}
   \date{Submitted: XXX; Accepted: XXX}

 \abstract{We study the appearance and specific properties of the structures in the local Universe by means of the Vlasov kinetic technique. We consider the role of the cosmological constant in local structure formation via the theorem on the general function that satisfies the identity of the gravity of the sphere and of the point mass. Then, the Hubble tension is naturally explained as a result of two flows, a local and a global one, with non-coinciding Hubble parameters. The linearized Vlasov-Poisson equation with the cosmological term is shown to lead to van Kampen's waves, Landau damping, and then to aperiodic structures. Aperiodicity thereby emerges as a intrinsic feature of the filamentary and void structure of the local Universe, and reveals the self-consistent field mechanism of its formation. The damping of the aperiodicity is then predicted and can be observationally traced upon the increase in the scale of the filaments. 
}


   \authorrunning{V.G. Gurzadyan, N.N.Fimin, V.M.Chechetkin}
   \titlerunning{On the origin of cosmological multi-connected structures}
   \maketitle
%
%________________________________________________________________

\section{Introduction}

The cosmological tensions that have recently emerged, most notably the Hubble tension \citep{R,R1,R2,R3,Dai,LR} and the baryon acoustic oscillation signature by
DESI DR2 \citep{DESI1,DESI2}, outline the possible genuine differences in the description of the early and late Universe. The theoretical approaches that address the tensions span a broad spectrum of issues, from the need for new physics to particular models of evolving dark energy and modified gravity, amongst other issues \citep{Cap,Alf,Chaud}. 
The origin and the evolution of large-scale low-dimensional structures, such as the long-known void walls and filaments of various scales, are among the goals of the theoretical approaches. The Zeldovich pancake theory \citep{Z,Arn,SZ} describes the evolution of the primordial density perturbations in hydrodynamical approximation and predicts the formation of the cosmic web on cosmological scales.    

The formation mechanisms of the filaments in the early and late Universe can have specific differences that  distinguish their features on various scales. The role of the self-consistent interaction in the structure formation in the local Universe was considered in  \citep{GFC1,GFC2,GFC3,GFC4}. The principal aspect in those studies was the consideration of the role of repulsive interaction due to the cosmological constant in the local scales. The cosmological constant in the local scales emerges if based on a theorem \citep{G1} that states the most general function for the force satisfying the identity of the gravitational field of a sphere and of a point mass, i.e. when the sphere and the point mass have an identical influence on a test particle. That general function has the form 
\citep{G1}
\begin{equation}
F=-{\frac{GMm}{r^{2}}}+{\frac{\Lambda c^{2}mr}{3}},
\end{equation}
where the second term in the right-hand side marks the cosmological constant term in weak-field General Relativity and McCrea-Milne non-relativistic cosmology \citep{MM,Z81}. That second term (i.e., the cosmological term) does not change the O(4) symmetry of the Newtonian
field. It is notable that the general function does not contain any other terms besides the second, the cosmological term. Also note that this function does not satisfy a force-free condition inside a spherical shell as distinct from Newtonian gravity. On this point, we mention the observational indications on the influence of galactic halos on the properties of the disks of spiral galaxies \citep{Kr}, which supports the presence of a force field inside a shell as predicted by Eq.(1). 
Then, Eq.(1) enables us to describe the Hubble tension as a result of two flows with non-identical Hubble parameters: a local flow and a global flow. The following two equations describe the two flows \citep{GS7,GS8}:
\begin{equation}
H_{local}^2 = \frac{8 \pi G \rho_{local}}{3} + \frac{\Lambda c^2}{3},
\end{equation}
\begin{equation}
H_{global}^2 = \frac{8 \pi G \rho_{global}}{3} + \frac{\Lambda c^2}{3}.
\end{equation}
The first equation follows from Eq.(1) and includes the local mean density, $\rho_{local}$ , as a parameter; the second equation is the Friedmann equation with the global mean density, $\rho_{global}$. The difference in the densities leads to the difference in the relevant Hubble parameters and thus provides an explanation for the Hubble tension. Moreover, as shown in \citep{GS8}, one can derive absolute constraints on the lower and upper values for the local Hubble parameter:
$$
\sqrt{\Lambda c^2/3} \simeq 56.2 < H_{local} < \sqrt{\Lambda c^2} \simeq 97.3 \,\,(km/sec\,\, Mpc^{-1}),
$$ 
which is in agreement with the observational data. This can be considered as an empirical support to the validity of the nonrelativistic description of the local Universe, as outlined in \citep{Z81}.
As shown in \citep{GS2,G2,GS3,GS4}, Eq.(1) fits the observational data on the dynamics of galaxy pairs, groups, and clusters.
We mention the recent efforts to test the law of Eq.(1) by means of quantum technologies \citep{PRR}.   

Although Debye screening is absent in gravitational systems and their problems have to be considered with different methods (e.g., \cite{GS86}), plasma theory has a well-developed and efficient mathematical apparatus for analyzing wave motions of various types that are suitable for adaptation to gravitational systems. 
The methods developed in plasma physics have been applied for certain problems of stellar dynamics \citep{Lynden1,Lynden2}, including those associated with Landau damping for small perturbations and with Bernstein--Greene--Kruskal waves \citep{Polyachenko1, Saslaw, Palmer, Vandervoort1,Polyachenko2}. In \citep{Lau1,Lau2} the authors point out the possibility of using van Kampen wave methods for the large-scale motion of clusters and galaxies. 

In this paper, we aim to describe large-scale structures using non-dissipative solutions of the Vlasov--Poisson equations of the van Kampen wave type. The periodicity of the waves is violated when taking into account the repulsive force due to the inclusion of a cosmological term in the consideration, since in this case our system is locally close to weakly inhomogeneous (for a long-range order, it is significantly inhomogeneous). Including the influence of the $\Lambda$--term
in the modified Poisson equation in the analysis of the behavior of a N-particle system follows from the above-mentioned theorem \citep{G1} and Eq.(1).
We also consider the possibility of introducing Bernstein--Greene--Kruskal waves \citep{Bernstein1,Montgomery2} as structural units of cosmological systems. For substantially inhomogeneous systems, we analyze the possibility of a smooth transition if we account for the field of gravitational disturbances to the normal mode method. Thus, aperiodicity emerges as a characteristic feature of the local filaments.

\section{Kinetic equations: Linearization}

We consider a set of $N$ cosmological objects, ``particles'' with masses $m_{i=1,...,N} =m\equiv 1$, which interact gravitationally.
Then the system of Vlasov--Poisson equations for description of its dynamics is represented as
$$
\frac{\partial F({\bf x},{\bf v},t)}{\partial t} +
{\nabla}_{\bf x}({\bf v}F)+\widehat{\mathcal{G}}(F;F)=0,
$$
\begin{equation}
\widehat{\mathcal{G}}(F; F) \equiv
-{\nabla}_{\bf v}F\cdot
{\nabla }_{\bf x}\Phi [F({\bf x})],
\label{1}
\end{equation}
\begin{equation}
{\Delta}_{\bf x}\Phi[F({\bf x})]=
4\pi A N \gamma \int F( {\bf x},{\bf v},t)\:d{\bf v}
- {c^2\Lambda},
\label{2}
\end{equation}
\noindent
where $F({\bf x},{\bf v},t)$ is the distribution function of gravitationally interacting
particles,
$A$ is a normalization factor for particle density, and $\gamma$ is the gravitational constant.
The system of particles is situated in a large domain of configurational space
$\Omega \subset {\mathbb R}^3_{\bf x}$ (${{\rm{diam}}\:\Omega \equiv R_\Omega}< \infty$).
The nonlinear Poisson Equation (\ref{2}) takes the form of an inhomogeneous Liouville--Gelfand equation
\citep{Du} with a local (kinetic) temperature \citep{V1}, when using a more general form for the Poisson equation.

Equation (\ref{2}) is the nonlinear Poisson equation for Newton--type gravitation.
The third term on the right hand side of the kinetic Equation (\ref{1}) may be represented as
\begin{equation}
\widehat{\mathcal{G}}(F; F) = {\bf G}\frac{\partial F}{\partial {\bf v}},~~~
{\bf G}\equiv -\nabla_{\bf x}\Phi[F({\bf x})],
\label{3}
\end{equation}
\begin{equation}
\Phi[F({\bf x},t)]= 4\pi AN \gamma \int\int {\mathfrak K}_3
({\bf x}-{\bf x}') F({\bf x}',{\bf v}',t)\:d{\bf x}'d{\bf v}'
\end{equation}
$$
+\frac{\Lambda c^2}{6}|{\bf x}|^2
+ \widehat{\mathfrak B}_{\partial \Omega} ({\bf x},{\bf x}'),
\label{4}
$$
where ${\mathfrak K}_3 ({\bf x}-{\bf x}')=-{|{\bf x}-{\bf x}'|^{-1}}$ (Newtonian interaction kernel) and
$\widehat{\mathfrak B}_{\partial \Omega} ({\bf x},{\bf x}')$
is an operator term that takes into account the influence of the boundary conditions (we take into account the influence of this
term by setting the appropriate boundary conditions).
Classical Newtonian potential $\Phi_{N}(r)= -\gamma M/r$ increases monotonically
on the interval $r\in (0,+ \infty )$ ($\Phi_N \in (-\infty,0)$), while the potential of Eq.(1),
including a cosmological term  $\Phi_{GN}(r)\equiv -{G M}/{r} - c^2\Lambda r^2/6$,
increases on the interval $r \in(0;\:r_c]$ and decreases on the interval $r \in(r_c;\:\infty)$, where $r_c = \big(3G M/(\Lambda c^2)\big)^{1/3}$.

We consider the nonstationary case of dynamics $F= F( {\bf x},{\bf v},t)$.
In previous publications \citep{GFC1,GFC2,GFC3,GFC4}  we focused on the possibility of transition to the integral form of the equation and the formulation of a boundary value problem
of the Dirichlet type (for the gravitational potential) with the aim of determining the Green's function of the problem for substitution into the kernel of the Hammerstein operator. However, for a nonstationary system of equations for the evolution of a cosmological system of particles in the self-consistent approximation (\ref{1})--(\ref{2}), the main role is played by
the formulation of the initial problem for the Vlasov equation. 
In this case, a direct derivation of the solutions and their study via analytical methods is complicated \citep{Camm2}. In the present work, we restrict ourselves to the study of the properties of the solutions
of the linearized version of the Vlasov--Poisson system for gravity for the potential of Eq.(1).

The linearization of the Vlasov equation is quite nontrivial, since its result depends significantly on the type of gravitational field and, due to the self-consistency of the problem, this depends on the distribution function of particles in the system.
From a physical point of view, it is natural to single out a stationary homogeneous solution when the distribution function does not depend on the coordinates $F=F_M({\bf v};T)$
or, in a more general case, $F=F_0({\bf v})$,
$F_0 \in C^1\cap L^2(\Omega_{\bf v}), \Omega_{\bf v}\subset {\mathbb R^3}_{\bf v}$;
it corresponds to the point at which the total force acting on the particle is zero, that is, the total potential of the gravitational attractive and repulsive forces is constant. Within the framework of the theorem \citep{G1}, one can take into account the presence of the previously mentioned local maximum of the two-particle potential. Then, the region near the equilibrium point, denote it ${\bf x}_0$, in the interaction channel between two distant external masses of subsystems includes the complete system of gravitating particles (we have a state of unstable equilibrium).

For a broader class of problems, it becomes necessary to consider a more general type of linearization near the equilibrium Maxwell--Boltzmann solution of the stationary Vlasov--Poisson system in the form
$F_{MB}\sim \exp\big( -{\mathfrak E} ({\bf x},{\bf v},t)/T \big)$,
including the (two--particle) potential, ${\mathfrak E}=m{\bf v}^2/2 + \Phi({\bf x},t_0)$, for a fixed instant of time. The dual solutions of the Poisson equation
$\Phi({\bf x},t_0)$ and the gravitational field strength are expressed through solutions of the Volterra equations of the second kind, and therefore classical dispersion relations for the Vlasov equations cannot be obtained.

It is necessary to note the meaning of the temperature, $T$ (kinetic temperature), in the solutions of the kinetic equation,
taking into account the action of the $\Lambda$--term in the Poisson equation.
Particle density on the right side of the Poisson equation can be expressed in terms of the nonstationary solution of the Vlasov equations.
In the simplest case, this solution
is identical to unimodal Maxwell distributions; in the general case, one can consider, for example,
representing $F_0$ as a multimodal set of Maxwellians with different amplitudes. However, the physical meaning of the equilibrium (nonuniform stationary states) solution of the Vlasov equation
is essentially different from that of the Boltzmann equation. This solution must meet the following requirements:
1)\:the maximum possible statistical independence, 2)\:isotropy of the velocity distribution, 3)\:stationarity of
distribution in the form $F({\bf x},{\bf v})=\rho({\bf x})\prod_{i=1,2,3}{\mathfrak f}(v_i^2)$.
The substitution of this expression into the Vlasov equation gives
\begin{equation}
\sum_i\bigg( v_i\frac{\partial \ln(\rho)}{\partial x_i} - \frac{\partial \Phi}{m\partial x_i}
\frac{\partial {\mathfrak f}(v_i^2)}{{\mathfrak f}(v_i^2) \:\partial v_i} \bigg)F=0,
\label{5}
\end{equation}
and we get a system of ordinary differential equations (ODE)
\begin{equation}
\frac{\partial(\ln\:\rho)/\partial x_i}{-\partial \Phi/\partial x_i} = \frac{\partial \ln\big({\mathfrak f}(v_i^2)\big/\partial v_i)}{m v_i}= -T^{-1},
\label{6}
\end{equation}
\noindent
where $T$ is a constant of separation of the variables; its physical meaning is kinetic temperature in the
system of interacting collisionless particles in accordance with Vlasov's definition \citep{V1,V2}, as
collisional equilibrium is globally absent in this system.

Equation (\ref{2}) for gravitational potential can be written as
\begin{equation}
\Delta \Phi({\bf x}) = \lambda^\dag \exp\big(-\Phi ({\bf x})/T \big)
-{c^2\Lambda},
~~~ \lambda^\dag =4\pi \gamma N A_T,~~
\label{7}
\end{equation}
$$
A_T\equiv \rho_0 \int \exp \big(-mv^2/(2T) \big)v^2\:dv,~~~\rho_0= \bigg( \frac{m}{2\pi T} \bigg)^{3/2}.
$$
The last equation can be rewritten in the form
$$
\Delta W({\bf x}) = {\lambda}^\sharp
\exp\big(W({\bf x})\big),
~~~ W\big({\bf x}\big)\equiv -\frac{\Phi ({\bf x})}{T}-\frac{c^2\Lambda {\bf x}^2}{6T},
$$
\begin{equation}
{\lambda}^\sharp =- \frac{\lambda^\dag}{T}\exp\big( c^2 \Lambda {\bf x}^2/T\big).
\label{8}
\end{equation}

Solutions of the equation $\Delta W({\bf x}) = -\zeta \exp\big(W({\bf x})\big)$ ($\zeta \in {\mathbb R}^1_{+}$) in the $3$--dimensional case
are radially symmetric ($W=W(|{\bf x}|)$ by the Gi--Nidas--Nirenberg theorem \citep{Du}) and
are unstable with respect to the pre-exponential parameter: their existence and number depend on the value of the parameter $\zeta$.
According to \citep{Bebernes}, the solution of the standard Dirichlet problem for it has a structure that can be described as follows.
Let $\zeta_{crit} = 2$ (if the boundary value problem is considered on the reduced interval
$|{\bf x}|\equiv r \in [0;1]$); then we have $\zeta_{FK} > \zeta_{crit}$ such that:
1)\:for $\zeta = \zeta_{FK}$, there is a unique solution ($W_{FK}$);
2)\:for $\zeta > \zeta_{FK}$, there are no solutions;
3)\:for $\zeta = \zeta_{crit}$, there is a countable infinity of solutions ($W_{crit}^{(n)}$, $n \in {\mathfrak V}$, $card({\mathfrak V})=\aleph_0$);
4)\:for $\zeta \in (0, \zeta_{FK})\backslash \{ \zeta_{crit} \}$, there is a finite number of solutions ($W^{(k)}_K$, $k\in \{1,2,..., K\}, K\ge 1$).
Since it is possible to uniquely (for fixed parameters $T,N$) compare the values of the function $-{\lambda}^\sharp (|{\bf x}|)$ with the values of the parameter $\zeta$, it can be stated that with an increase in the modulus of the radius vector ${|{\bf x}|}$, three regions of solutions to Equation (\ref{8}) arise:
the region of uniqueness of solutions $X_1({\bf x})=\{|{\bf x}|<X^{(I)}\}\cup \{X^{(III)}\}$, the region of multivalued solutions (differing in norm) $X_2({\bf x})=\{X^{(I)}<|{\bf x}|<X^{(II)}\}$,
and the region of absence of solutions $X_3({\bf x})=\{|{\bf x}|>X^{(III)}\}$.

Let us consider the linearization of Equation (\ref{7}) in the neighborhood of the solution $W({\bf x})$
analytic solution to which $W$ can be associated to solution (\ref{7}) as
$W+{\mathfrak w}({\bf x}, t)$ ($\|{\mathfrak w}\|\ll \|W\|$ by the chosen norm and,
correspondingly, $\exp({\mathfrak w})\approx 1+{\mathfrak w}$). We obtain the linear Poisson equation
\begin{equation}
\Delta {\mathfrak w} ({\bf x})={\lambda^\sharp}\exp\big(W({\bf x}) \big)\cdot {\mathfrak w}({\bf x}).
\label{9}
\end{equation}
Obviously, in the neighborhood of ${\bf x}_0$ the last equation is simplified, since the gravitational field of a point with an equivalent total
mass and cosmological repulsion allows us to set $\Phi({\bf x}_0)\equiv -T W({\bf x}_0) =\Phi^{(0)} (={\rm{const}})$.
Thus, the equation for the potential perturbation in the above neighborhood $O({\bf x}_0)$ takes the form ($K^{(0)}=\exp\big(W({\bf x}_0) \big)$)
\begin{equation}
\Delta {\mathfrak w} = {\lambda^\sharp} ({\bf x}_0)K^{(0)} \cdot {\mathfrak w}({\bf x}).
\label{10}
\end{equation}
The linearization of the Vlasov equation itself is performed (in the simplest case considered)
in the neighborhood of the equilibrium function $F_M({\bf v})$ or, in a more general case, $F_0({\bf v})$ with several maxima,
which is realized, for example, in the case of co-directional particle beams; so we have
\begin{equation}
F\:\rightarrow\: F_{MB}({\bf x},{\bf v})+\tilde{f}({\bf x},{\bf v},t),~~~
\label{11}
\end{equation}
\noindent
where the perturbation $\tilde{f}({\bf x},{\bf v},t)$ is related to the Poisson Equation (\ref{9}) with an exponential dependence of the parameter on the spatial variable.
Eliminating the quadratic terms a small addition $\tilde{f}$ gives us
\begin{equation}
\frac{\partial \tilde{f}}{\partial t}+ {\bf v}\cdot\nabla_{\bf x}\tilde{f} -
\nabla_{\bf v}F_{0}\cdot \nabla_{\bf x} \phi[\tilde{f}]({\bf x},t)=0,~~~-T \big(W+{\mathfrak w}\big)=\Phi+\phi.
\label{12}
\end{equation}

Next, we consider the methodology for studying the linear system of Vlasov-Poisson equations using ``normal modes'' and the use of the transition to the space of distributions. This allows us to study analogs of the attenuation of Landau waves and longitudinal van Kampen density waves for a system of gravitating particles.

\section{Van Kampen modes versus self-consistent gravitational potential with a cosmological constant}

Let us consider the invariant properties (independent of solutions) of the linearized Vlasov--Poisson system of Equations (\ref{10})--(\ref{11}). First, we consider the case of a gravitational field strength that corresponds to a local neighborhood of the extremum of the self-consistent potential, taking into account the action of the cosmological term
\begin{equation}
\frac{\partial {\tilde{f}}({\bf x},{\bf v},t)}{\partial t} +{\bf v}\nabla {\tilde{f}}({\bf x},{\bf v},t)=
\nabla_{\bf x} {{\phi}}({\bf x},t) \cdot \nabla_{\bf v} F_0 ({\bf v}),~~~
\label{13}
\end{equation}
\[
\nabla_{\bf x} {\phi}({\bf x},t)=
{\lambda^\sharp}_0 K^{(0)} \int_{{\Omega_{{\bf x}'}}}
\int_{\Omega_{\bf v}}\nabla_{\bf x} \frac{{\tilde{f}}({\bf x}',{\bf v},t)}{|{\bf x}-{\bf x}'|} d{\bf v} d{\bf x}'.
\]

We represent $\tilde{f}({\bf x}, {\bf v},t)$ via the van Kampen ansatz or ``normal modes'' \citep{vankampen1,vankampen3}:
${\mathfrak{K}}({\bf v})\exp(i{\bf k}{\bf x}-i \omega t)$
(plane waves are eigenfunctions of the Laplacian from the left-hand side of the Poisson Equation (\ref{10})).
We are interested in solutions of the system of Equations (\ref{10})--(\ref{11}) in the form of longitudinal waves, therefore in the velocity space
we choose axes that are parallel ($z$) and perpendicular ($x,y$) to the wave vector, ${\bf k}$; then the longitudinal component of the velocity is
$v_{\|}=v_z={\bf e}_{\bf k}\cdot{\bf v}$ (where ${\bf e}_{\bf k}={\bf k}/|{\bf k}|$),
and the transverse component is, respectively: ${\bf u}={\bf v}-{\bf e}_{\bf k}v_{\|}$. In this case, we can introduce distribution functions that only depend on one component of the velocity: ${f}({\bf k}, v_{\|},t)=\int \tilde{f}({\bf k}, {\bf v},t)\delta(v_{\|}-{\bf k}\cdot{\bf v}/k)d{\bf v}
=\int \tilde{f}({\bf k},{\bf v},t)d{\bf u}$.

We rewrite Equation (\ref{11}) for these modes, freeing ourselves from the transverse velocity components (and discarding the tilde sign over $f$):
\begin{equation}
\big(\omega - kv_{\|}\big)\int{\mathfrak{K}}({\bf v})d{\bf u}= \frac{4\pi}{k^2}{\bf k}{\lambda^\sharp_0}K_\Lambda^{(0)}\int\frac{\partial F_0}{\partial v_{\|}}d{\bf u}
\int {\mathfrak{K}}({\bf v}')d{\bf v}',
\label{14}
\end{equation}
$$
-\widehat{\phi}({\bf k},t)={\lambda^\sharp_0} K_\Lambda^{(0)}
\int\int \frac{{\mathfrak{K}}({\bf v}')({\bf x}-{\bf x}')}{|{\bf x}-{\bf x}'|^3}\exp(i{\bf k}{\bf x}'-i \omega t) d{\bf v}'d{\bf x}',
$$
$$
\int \frac{\bf x}{|{\bf x}|^3}\exp(i{\bf k}{\bf x})d{\bf x}= 4\pi i \frac{\bf k}{k^2}.
$$

We divide both sides of the last equation by $\big(\omega - kv_{\|}\big)$ and integrate with respect to the variable $v_{\|}$. The integral $\int {\mathfrak{K}}({\bf v}')d{\bf v}'$ (an unimportant constant) is canceled out,
and we obtain a dispersion relation that is invariant with respect to the form of the solution of the kinetic equation
\begin{equation}
1- (\varkappa/{k}) \int \frac{df_0}{d v_{\|}} \frac{d v_{\|}}{\omega - kv_{\|}} =0,
~~~\varkappa = {4\pi}\lambda^\sharp_0 K_\Lambda^{(0)}.
\label{15}
\end{equation}
If we do not consider the longitudinal velocity as distinguished, then the general form of the dispersion law has the form: $D({\bf k},\omega)\equiv 1-\varkappa({\bf k}/k^2) \int_{L} (F_0)_{\bf v}'(\omega - {\bf k}{\bf v})^{-1} d{\bf v}=0$ (normal modes will
correspond to the case ${\rm Re}(\omega(k))\gg {\rm Im}(\omega(k))$).
 
We are interested in the possibility of obtaining a solution to the Vlasov--Poisson equations that is stable in time and associated with the simplest cosmological structures, i.e. those of low dimensionality. It can be obtained using normal modes in the form
\begin{equation}
f(z,v_{\|},t)=\int\int {\mathfrak{O}}(k,\nu){\mathfrak{N}}(k,\nu; v_{\|})\exp\big( ikz-ik\nu t \big)\big|_{\nu=\omega/k}dk d\nu,
\label{16}
\end{equation}
\noindent
where ${\mathfrak{O}}(k,\nu)$ is an (admissible) function that corresponds to certain Cauchy data for the kinetic equation for perturbation $f$.
If the initial condition is represented as $f(z,v_{\|},t=0)=\int g(k,v_{\|})\exp(ikz)dk$, then, obviously,
Equation (\ref{16}) is reduced to the form
$\int {\mathfrak{O}}(k,\nu){\mathfrak{N}}(k,\nu; v_{\|})d\nu = g(k,v_{\|})$, and the variable $k$ here acquires the meaning of a parameter.

For what follows, we return to Equation (\ref{14}) and consider a nonobvious consequence of taking the integral of ${\mathfrak{K}}$ over the transverse velocities and dividing both parts by $\big(\omega - kv_{\|}\big)$. The result here must take into account the possibility of the equation solutions going into the space
of generalized functions: as is known, for the functional equation $(x-y)\mu_1 (x)=\mu_2 (x)$ (defined on the interval $[x_1;x_2]$ of the real axis) and the point $y\in (x_1;x_2)$,
the solution must be interpreted as a distribution. This distribution can be written in the following form:
$\mu_1 (x|y)=\mu_2 (x)P.V.\frac{1}{x-y} + \mu^\natural(y)\delta (x-y)$, where the Cauchy principal value in the form of a distribution is defined by the relation
$(P.V.\frac{1}{x},\mu) = \lim_{\epsilon\to 0}\int_{|x|\ge \epsilon} (\mu(x)/x)dx$), and $\mu^\natural(y)$ is ``the strength of the concentration'' of the Dirac function at the point $x=y$
determined from additional conditions imposed on the generalized function $\mu_1 (x|y)$.

Thus, Equation (\ref{14}) rewritten as
\begin{equation}
(\nu - v_{\|}){\mathfrak N}(v_{\|})=\nu_\varkappa{\mathfrak F}(v_{\|})\int {\mathfrak{K}}(v_{\|}')dv_{\|}',\:\:~~\int {\mathfrak{K}}(v_{\|}')dv_{\|}'=1,
\label{17}
\end{equation}
$$
{\mathfrak N}(v_{\|})\equiv \int {\mathfrak{K}}({\bf v})d{\bf u},
$$
$$
{\mathfrak F}(v_{\|})\equiv \int\frac{\partial F_0({\bf v})}{\partial v_{\|}}d{\bf u},~~~\nu=\frac{\omega}{k},~~~\nu_\varkappa=\frac{\varkappa}{k^2},~~~
$$
\noindent
after dividing both sides of Equation (\ref{17}) by $(\nu - v_{\|})$ and should be written in the sense of distributions
\begin{equation}
{\mathfrak N}(v_{\|})=\nu_\varkappa \cdot P.V.\frac{{\mathfrak F}(v_{\|})}{\nu - v_{\|}}+\mu^\natural\delta(\nu - v_{\|}),~~~(\nu - v_{\|})\delta(\nu - v_{\|})=0,
\label{18}
\end{equation}
\noindent
In this case, from the normalization condition in Eq. (\ref{17}), the intensity value $\mu^\natural$ is determined by the condition of its agreement with Formula (\ref{18}):
$\mu^\natural=1-\nu_\varkappa P.V.\int \big({{\mathfrak F}(v_{\|})}/(\nu - v_{\|})\big)dv_{\|}$.

Let us substitute
into the equation $\int {\mathfrak{O}}(k,\nu){\mathfrak{N}}(k,\nu; v_{\|})d\nu = g(k,v_{\|})$ the value ${\mathfrak N}(v_{\|})$ from (\ref{18}):
\begin{equation}
{\mathfrak{O}}(k,v_{\|})\big( 1-\pi \nu^2_\varkappa \widehat{\mathfrak{H}}{\mathcal F}(v_{\|}) \big)-
\widehat{\mathfrak{H}}{\mathfrak{O}}(k,v_{\|})\pi \nu^2_\varkappa {\mathcal F}(v_{\|}) = g(k,v_{\|})
\label{19}
\end{equation}
\noindent
($k$ is still a parameter). Here, $\widehat{\mathfrak{H}}(\Psi(x))=(1/\pi)P.V.\int \big(\Psi(x')/(x-x')\big)dx'$ is the Hilbert transform,
which is related to the Fourier transform of the function $\Psi(x)=\Psi_{+}(x)+\Psi_{-}(x)$: $\Psi_{+}(x)-\Psi_{-}(x)=i \widehat{\mathfrak{H}}(\Psi(x))$,
where $Y_{+}(x)\equiv \int^\infty_0 Y(q)\exp(iqx)dq$, $Y_{-}(x)\equiv \int_{-\infty}^0 Y(q)\exp(iqx)dq$; symbols $Y,Y_{\pm}$ are used to denote
functions $\Psi,\mathcal F,{\mathfrak{O}},g,$ and their decompositions.

The last equation can be rewritten as
\begin{equation}
\big( 1+2\pi i \nu^2_\varkappa {\mathcal F}_{+}(v_{\|}) \big) {\mathfrak{O}}_{+}(k,v_{\|}) +
\big( 1-2\pi i \nu^2_\varkappa {\mathcal F}_{-}(v_{\|}) \big) {\mathfrak{O}}_{-}(k,v_{\|}) =
\end{equation}
$$
 g_{+}(k,v_{\|})+ g_{-}(k,v_{\|})\equiv g (k,v_{\|}).
\label{20}
$$
The terms on the left-hand side are analytic and have no singularities in the upper ($Im(\eta)>0$) and lower ($Im(\eta)<0$) parts of the complex
$(\eta_{Re},\eta_{Im})$--plane (${\mathbb R}\ni v_{\|}\to\eta \in {\mathbb C}$), respectively,
and also asymptotically tend to zero in their half-plane. The decomposition of $g(\eta)$ into two functions with such properties is unique, and therefore
$\big( 1\pm2\pi i \nu^2_\varkappa {\mathcal F}_{+}(v_{\|}) \big) {\mathfrak{O}}_{\pm}(k,v_{\|})= g_{\pm}$. Therefore, if there is a solution to Eq.(\ref{19}),
then it must coincide with ${\mathfrak{O}}={\mathfrak{O}}_{+}+{\mathfrak{O}}_{-}$, ${\mathfrak{O}}_{\pm}=g_{\pm}/(1+2\pi i \nu^2_\varkappa {\mathcal F}_{\pm})$ (the condition
for this is ${\mathfrak{N}}(v_{\|})\neq 0$, which is true, in particular, for the Maxwellian distribution). If we consider on the half-plane $Im(\eta)>0$ a holomorphic and asymptotically close to unity
function ${\mathcal Z}(\eta)=1+ 2\pi i \nu^2_\varkappa {\mathcal F}_{+}(v_{\|})$, we can extend it to the half-plane $Im(\eta)<0$:
${\mathcal Z}(\eta)=1+4\pi^2 i \nu^2_\varkappa z {\mathfrak{N}}(\eta)+2\pi \nu^2_\varkappa \int \eta' {\mathfrak{N}}(\eta')/(\eta'-\eta)d\eta'$.
Now we can write out the final form of the solution to the initial value problem with the general solution (\ref{16}):
$$
f(z,v_{\|},t)= (2\pi)^{-1}\int\int\int {\mathfrak{N}}(k,\nu; v_{\|})\exp\big( ik(z-z')-ik\nu t \big)
$$
$$
\big( f_{+}(z',\nu,t=0)/{\mathcal Z}(k,\nu) +
\label{21}
$$
$$
+ f_{-}(z',\nu,t=0)/\overline{{\mathcal Z}(k,\nu)} \big) dk dz' d\nu,
$$
\begin{equation}
f_{+}(z,\nu,0) +f_{-}(z,\nu,0)=f(z,\nu,0).
\end{equation}

For the initial function of the form $f(z,v_{\|},t=0)=\int g(v_{\|})\exp(ik z)\delta(k-k_1)dk$ ($\lambda=2\pi/k_1={\rm const}$), the density  of particles in the disturbance wave is
$$
\varrho_f(z,t)=\exp(ik_1 z)\int_{\mathbb R}\exp(-ik_1 \nu t)\big( g_+(v_{\|})/{\mathcal Z}(k_1,\nu)+
$$
$$
g_{-}(v_{\|})/\overline{{\mathcal Z}(k_1,\nu)}d\nu.
$$

In this case, since $g_{-}(\nu)$ is defined through negative frequencies, and $\overline{{\mathcal Z}(\nu)}$
is holomorphic in the lower half-plane and is bounded by unity at infinity, then the integral of $g_{-}/\overline{{\mathcal Z}}$ tends to zero as $t>0$.
Therefore, $\varrho_f(z,t)=\int \exp(ik_1 z-ik_1 v_{\|}t)\big(g_{+}(v_{\|})/\overline{{\mathcal Z}}(v_{\|}) \big)dv_{\|}$.

Assuming that ${\mathcal Z}(\nu)$ can be continued analytically into the strip $Im(\nu)\in [-|\nu_{min}|; 0]$, and there exists a quantity $\nu_0=\nu^\dag - i \nu^\dag_{*}$
($\nu^\dag \in {\mathbb R}$, $\nu^\dag_{*} \in (0,|\nu_{min}|)$), we can shift the integration path $\int_{\mathbb R}$ on the left-hand side of the expression for $\varrho_f(z,t)$
parallel to the real axis down, below the point $\nu_0$: $Im(\nu)=-\nu_{Im}$, $\nu_{Im} \in (\nu^\dag_{*},|\nu_{min}|)$. The contribution to the integral from this pole can be obtained by the residue theorem: $\varrho_f(z,t)=-2\pi i \exp(ik_1 z-i k_1 \nu_0 t)\big( g_{+}(\nu)/{\mathcal Z}_{\nu}'(k_1,\nu)\big|_{\nu=\nu_0}$.
Since $ik_1 \nu_0 t= ik_1 \nu^\dag t+ik_1 (-i \nu^\dag_{*}) t$, the described density wave will be damped with a real damping coefficient $\beta=k_1 \nu^\dag_{*}$
($\beta^{-1}$ --- the wave decay time); that is, in the lower region of the complex plane, Landau damping \citep{Landau,Maslov1} is observed.
To determine $\nu^\dag_{*}$ and $\nu_{Im}$, we use the expansion of the function $Z$ in the neighborhood of the point $\nu^\dag$: $Z(\nu^\dag)- i \nu^\dag_{*}(dZ/d\nu)(\nu^\dag)=0$.
Thus, if we isolate the real part of the equation (${Re(Z)}(\nu^\dag)=0$), we determine the condition on the phase velocity
$\nu^\dag$: $P.V.\int \nu {f_0}(\nu)/(\nu^\dag-\nu)d\nu=(2\pi \varkappa/k_1^2)^{-1}$; if we isolate the condition on the imaginary part, we obtain:
$\pi \nu^\dag {f_0}(\nu^\dag)=\nu^\dag_{*} P.V. \int\nu {F_0}(\nu)/(\nu^\dag - \nu)^2 d\nu$.

Thus, we obtain a complete description for the density waves of self-gravitating particles moving in one direction, provided that the potential perturbations
in the neighborhood of its macro-extrema point (for the equilibrium function $F_0({\bf v})$, coinciding with or being a direct generalization of the Maxwellian)
obey the linearized Poisson equation.
Van Kampen waves admit a more general form of the ansatz, when normal modes have a more universal form than plane waves \citep{Case1}. We demonstrate its application
to the system of gravitating particles under consideration, which is essential for the two-dimensional geometry of a system with rotation.

Consider the ``conjugate'' problem to (\ref{17}) in the following form:
$$
(\nu - v_{\|}){\mathfrak A}({\bf k},v_{\|};\omega^\ddag)=\int\nu_\varkappa (k,v){\mathfrak A}({\bf k},v;\omega^\ddag)dv,
$$
\begin{equation}
\int \nu_\varkappa (k,v){\mathfrak A}({\bf k},v;\omega^\ddag)dv=1,
\label{22}
\end{equation}
$$
(\nu^\ddag - v_{\|}){\mathfrak A}({\bf k},v_{\|};\omega^\ddag)=1,~~~(\nu^\ddag - \omega^\ddag)f({\bf k},v_{\|};\omega^\ddag){\mathfrak A}({\bf k},v_{\|};\nu^\ddag)=0,
$$
\noindent
where normal modes are introduced by the relation ${\mathfrak A}({\bf k},v_{\|},t)={\mathfrak A}({\bf k},v_{\|};\omega)\exp(-i\omega t)$.
If the real eigenvalues $\omega^\ddag$ are not zeros of the function $\nu_\varkappa (k,v)$, then the eigenfunctions that correspond to them take the form ${\mathfrak A}({\bf k},v_{\|};\omega^\ddag)= \nu^\ddag (k,\omega^\ddag)\delta(\omega^\ddag - v_{\|})+P.V.\big(1/(\omega^\ddag - v_{\|})\big)$; further,
we should consider the cases when: 1)\:$\omega^\ddag$ are zeros of the function $\nu_\varkappa (k,v_{\|})$, but not $\nu^\ddag (k,v_{\|})$;
2)\:$\omega^\ddag$ are the zeros of the functions $\nu_\varkappa (k,v_{\|})$ and $\nu^\ddag (k,v_{\|})$; 3)\:$\omega^\ddag_j$ are the complex zeros of ${\mathfrak A}({\bf k},v_{\|};\omega^\ddag_j)$.
Finally, we obtain
$$
{\mathfrak A}({\bf k},v_{\|};\omega^\ddag)=\sum_j C(k,j){\mathfrak A}({\bf k},v_{\|};\omega^\ddag_j)+
$$
$$
\int C(k,j)\omega^\ddag
{\mathfrak A}({\bf k},v_{\|};\omega^\ddag)d\omega^\ddag.
$$
The amplitude of the modes is obtained as the sum over the discrete and continuous spectra of the singularities of the functions $\nu_\varkappa (k,v_{\|})$ and $\nu^\ddag (k,v_{\|})$.

Thus, van Kampen waves in the linear approximation for the Poisson equation, with initial conditions that only depend on the particle velocities, can serve as a basis for the quasi-local approximation near the extremum point of the self-consistent potential. In the formulation of the problem of the evolution of cosmological structures, such an approach is applicable for the initial stages of the process of their formation, when the gravitational interaction does not yet have a significant effect on the topological properties of the selected system of particles. It would be interesting to estimate the
change in the sizes of proto-structures during the transition to the phase of gravitational interaction dominance
from the point of view of an absence of solutions to Equations (\ref{8}), since this would lead to the proto-structures to a quasi-Jeans-type decay caused by the presence of an additional
term -- the cosmological term -- in the Liouville-Gelfand equation.

\section{Aperiodic structures}

In addition to van Kampen waves, the Vlasov--Poisson system of equations has wave solutions of a very general type, which can also be associated with cosmological structures.
We are talking about one-dimensional Bernstein--Green--Kruskal (BGK) waves \citep{Bernstein1,Montgomery2,Bernstein2}. For the simplest one-dimensional case, the Vlasov equation in coordinates
$({\mathfrak E},x,t)$ (${\mathfrak E}=mv^2/2 +m\Phi (x)$ is the energy of a particle in a gravitational field:
\begin{equation}
\frac{\partial F({\mathfrak E},x,t)}{\partial t} +v(x,{\mathfrak E})\frac{\partial F}{\partial x} + \big( v(x,{\mathfrak E})/m \big)
\big( G(x,t)-\Phi'(x) \big)\frac{\partial F}{\partial {\mathfrak E}}=0,
\label{23}
\end{equation}
$$
-\frac{\partial G}{\partial x}=4\pi\gamma N \int f({\mathfrak E},x,t)dv - c^2\Lambda
$$
where the second term on the right-hand side corresponds to the repulsive potential, as before.
At equilibrium, $f=f_0({\mathfrak E})$, ${\mathfrak E}= -d\Phi/dx$; if we set $F=F_0({\mathfrak E})+f(x,{\mathfrak E},t)$, $G(x,t)=-\Phi'(x)+ G_1(x,t)$,
then the linearized Vlasov equation takes the form
\begin{equation}
\big(v(x,{\mathfrak E})\big)^{-1}\frac{\partial f}{\partial t}+\frac{\partial f}{\partial x}-\frac{G_1}{m} \frac{dF_0}{d{\mathfrak E}}=0
\label{24}
\end{equation}
(the repulsive potential is absent in the equation for perturbations, since its effect is present in the basic macro-potential $\Phi (x)$).
We seek a solution to the equation in the form $f=\psi (x)\exp(-i\omega t)$:
$$
\frac{\partial \psi}{\partial x}-i\omega v^{-1}(x,{\mathfrak E})\psi=G_1/m \cdot \frac{dF_0}{d{\mathfrak E}},
$$
\begin{equation}
-i\omega G_1 =4\pi\gamma \int \psi v dv= 4\pi \gamma \int_{{\mathfrak E}_0}^\infty \psi (x,{\mathfrak E})d{\mathfrak E},
\label{25}
\end{equation}
\noindent
where $v^{-1}(x,{\mathfrak E})=(2{\mathfrak E} + \Phi)^{-1/2})$. If we exclude $G_1$ from the last two equations, we obtain an equation of the form
\begin{equation}
\frac{\partial \psi}{\partial x}-i\omega v^{-1}(x,{\mathfrak E})\psi= (4\pi i/m)\omega^{-1}\frac{dF_0}{d{\mathfrak E}}\int_{{\mathfrak E}_0}^\infty \psi\:d{\mathfrak E}.
\label{26}
\end{equation}
If $\Phi \to 0$, then the last equation coincides with the eigenvalue equations obtained in the van Kampen method. Therefore, following the previously considered
method, we select the ``normal''
mode with a fixed wave number $k=K_1$ and the corresponding frequency $\Omega^{(0)}$ related via the dispersion relation (\ref{26}):
\begin{equation}
ik \psi (k; K_1)-i\omega \int_{\mathbb R} v^{-1}(q) \psi(k-q; K_1)dq=
\end{equation}
$$
i(4\pi F_0' \gamma/m)/\omega \int_{{\mathfrak E}_0}^\infty \psi(k; K_1)d{\mathfrak E}.
\label{27}
$$
This equation can be solved by expanding in powers of the parameter $\Phi(k)/{\mathfrak E}_0$: $\Theta_j (k) =\sum_{k=0,...,\infty}\Theta^{(j)}(k)$, where
$\Theta{(j)}(k)\in \{v^{-1}(k),\psi (k;K_1), \omega$\}. Putting $v^{1}(k)=\delta (k)/\sqrt{2{\mathfrak E}}$,  in the zeroth approximation we obtain two types of eigenmodes,
discrete and continuous
$\psi^{(0)} (k;K_1) = \big( K_1 -\omega^{(0)}/\sqrt{2{\mathfrak E}}\big)^{-1} \big( (4\pi F_0' \gamma/m)/ \omega^{(0)} \big)\delta (k-K_1)$. The criterion for discreteness of the quantities $\omega^{(0)}$ are the conditions $(4\pi F_0' \gamma/m)\cdot \big( (\omega^{(0)})^2 /(2K_1^2) \big)= 0$, or
the condition ${Im}(\omega^{(0)})\neq 0$. If $(4\pi F_0' \gamma/m)\cdot \big( (\omega^{(0)})^2 /(2K_1^2) \big)\neq 0$,
the functions $\psi^{(0)}(k;K_1)$ should be considered in the class of distributions, since
$\big( K_1 -\omega^{(0)}/\sqrt{2{\mathfrak E}}\big)^{-1}=P.V. \big(1/\big( K_1 -\omega^{(0)}/\sqrt{2{\mathfrak E}}\big)\big) + (\mu^\ddag)^{(0)}(K_1,\omega^{(0)})
\delta( K_1 -\omega^{(0)}/\sqrt{2{\mathfrak E}})$ (in this case $Im(\omega^{(0)})\le 0$ indicates the asymptotic stability of the complete solution.
In a similar way, one can obtain $\psi^{(1,2,...)}(k;K_1)$.

The main result after constructing the appropriate number of terms in the series for $\psi (k;K_1)$ is the establishment of the density function of the solution of the BGK equations.
This expression can be used for comparative calculations of the macro-parameters of cosmological objects (see below).

As can be seen from the form of Equation (\ref{26}), the initial condition is also taken in the form of a (generalized) Maxwell function, and the methodology of further research makes significant use of this.
To what extent is it legitimate in general to use $F_{MB}$ in the role of the Cauchy conditions for the Vlasov equation for cosmological systems (for the linearized case, $f^{(0)}({\bf x},{\bf v})$)?
In accordance with the structure of Equation (\ref{12}), the formal substitution of normal modes (of the form ${\mathfrak K}_1({\bf v}){\mathfrak K}_2({\bf x},t)$, in
the simplest case ${\mathfrak K}_2(z,t)=\exp(ikz-i\omega t)$)
into this equation at $F|_{t=0}=F_{MB}({\bf x},{\bf v})$ will lead to the appearance of a bilinear dependence
on the spatial and temporal variables, which indicates a nonlocal form of interaction of carrier waves, which should be described by an integral relation, which
excludes the presence of a local differential dispersion formula. Apparently, the most
direct way to study the properties of the linear Vlasov equation for an inhomogeneous field and initial conditions
lies through finding the explicit form of the force interaction term (for $F_0\to F_{MB}$).

In this case, there are obviously problems
when substituting into the equation decomposition solutions of a priori form with independent modulation by coordinates of the extended phase space.
Following \citep{Maslov2,Maslov1}, we assume that the characteristics of the linear (complete) Vlasov equation coincide with the phase trajectories of the Hamiltonian system $d{\bf X}/dt=V$, $d{\bf V}/dt=-d\Phi/d{\bf X}$, since one should consider the additional term ${\mathcal T}(\Phi,f) \equiv -\Phi_{\bf x}'f_{\bf v}'$ on the
left-hand side of Equation (\ref{13}); the spatial changes in the potential of the ``main'' gravitational field of the system are taken into account.
The function $\Phi$  satisfies Equation (\ref{7}) (or (\ref{8}), if after obtaining the solution we pass from the dependent variable $W$ to $\Phi$). The solution of this dynamical system with the initial conditions
${\bf X}\big|_{t=0}={\bf x}$, ${\bf V}\big|_{t=0}={\bf v}$ is as follows:
${\bf X}({\bf x},{\bf v},t)$, ${\bf V}({\bf x},{\bf v},t)$ ($t\in {\bf R}^1$). The first integral of the dynamic system is ${\mathfrak E}=m{\bf v}^2/2+\Phi ({\bf x})$
(which corresponds to the conservation of energy along the trajectories of the Vlasov equation in the spatially inhomogeneous case, and this is why the term ${\mathcal T}(\Phi,f)$ was introduced). For the function $f({\bf x},{\bf v},t)$, through the shift along the trajectories from the initial point, we have the IInd type Volterra equation
$$
f({\bf x},{\bf v},t)=f^{(0)}\big({\bf X}({\bf x},{\bf v},-t),{\bf V}({\bf x},{\bf v},-t)\big) + 
$$
\begin{equation}
\frac{dF_0}{d{\mathfrak E}}\int^t_0
\nabla \phi \big( {\bf X}({\bf x},{\bf v},\xi-t),\:\xi \big) {\bf V} \big( {\bf x},{\bf v},\xi-t \big)\:d\xi,
\label{28}
\end{equation}
\noindent
and, after substituting this expression into the Poisson equation $\nabla^2 {\phi} = {\lambda^\sharp}\exp\big(W({\bf x}) \big)\cdot \phi({\bf x})$,
we have an explicit form for the force term (${\bf G}\to {\bf G}[\Phi]+{\bf g}[\phi]$ when linearized):
\begin{equation}
- {\mathfrak y}^{-1} {\nabla \phi}({\bf x},t) =\int \int f^{(0)}\big({\bf X}({\bf x},{\bf v},-t),{\bf V}({\bf x},{\bf v},-t)\big)d{\bf v}d{\bf x}+
\label{29}
\end{equation}
$$
+
\int\int\int^t_0 \frac{dF_0}{d{\mathfrak E}}
{\nabla\phi}\big( {\bf X}({\bf x},{\bf v},-\xi),\:t-\xi \big) {\bf V}\big({\bf x},{\bf v},-\xi \big)d\xi d{\bf v}d{\bf x},
$$
\noindent
where the notation ${\mathfrak y} \equiv \lambda^{\sharp}\exp\big(-\Phi({\bf x})/T \big)$.
In accordance with the definition in Formula (\ref{8}) for the potential value $W({\bf x}),$
for the motion in a nonuniform field of a system of gravitating
particles, we obtain  the influence of two integrand factors at once: ${dF_0}/{d{\mathfrak E}}\cdot{\bf g}$. This is due to the fact that
both ${\mathfrak E}$ and ${\bf g}$ contain the full Liouville--Gel'fand potential. This significantly complicates the consideration of the question of the uniqueness
of the solution, since the values of the potential $W({\bf x})$ in these factors may lie in different regions $X_i ({\bf x})$ from Section 2. Apparently, in order to
establish the uniqueness of the solution, the behavior of the function ${\bf v}({\bf x})\big|_{{\mathfrak E}=const}$ should be considered. In addition, the question arises of the physical
manifestation of the multi-valuedness of solutions to the Vlasov--Poisson equation in the region $X_2({\bf x})$: since the norms of the solutions $W({\bf x})$ with the same pre-exponential factor
differ by finite values, the standard definition of bifurcation of solutions is inapplicable, and smooth solutions of the Vlasov equation that correspond to the minimal norm of the solution must collapse.
However, ``destruction of the solution'' can be expressed in an increase in its norm, for example, due to an increase in the density of particles, which can be a time-dependent process.
Consequently, in addition to the wave form of motion, in the simplest case considered in Section 3 using the example of van Kampen waves,
there may be processes of local ``thickening'' of matter over time in a certain region of space (antinodes of a longitudinal wave, in particular), associated with the transition in the region of multivalued solutions of the Liouville-Gel'fand equation to a new norm of its solution.

We point out that the left-hand side of the Vlasov--Poisson equation with an additional term ${\mathcal T}(\Phi,f)$ as $\Phi\to {\rm const}$ tends can be assumed to be
extremely close to the ``classical'' left-hand side of the linearized Equation (\ref{13}), however, the right-hand side of the kinetic equation, containing
the second term of the right-hand side of the Volterra Equation (\ref{28}), will retain an unchanged form (${\mathfrak E}\approx {\bf v}^2/2 +\Phi({\bf x}_0)$),
and this part only slightly depends on function $f$. Consequently,
we can formally consider the representation of the solution in the form of a normal mode of the above-considered ``ansatz'' type, divide both parts by $(\omega - \nu)$ (taking into account the occurrence of the term in the form of a distribution), and repeat all the operations of Section 3. In this regard, van Kampen waves can also be used for the spatially--(weakly)inhomogeneous case.
Let us demonstrate this by turning to the one-dimensional case (that corresponds to the previously considered longitudinal waves) for the sake of clarity of the calculations. We integrate both
parts (\ref{29}) over the interval $[0, \tilde{z}]$, rearrange the order of integration, and make a change of the variables $\eta_X=X(x,v,-\xi)$, $\eta_V =V(x,v,-\xi)$ in the second term of the right-hand side.
Since ${\mathfrak E}(X,V)={\mathfrak E}(\eta_X,\eta_V)$, $dXdV=d\eta_X d\eta_V$, this term will take the form of a flow through the surface:
$\int\int_\sigma {\mathfrak E}(\eta_X,t-\xi) \partial \big(F_0(\eta_V^2/2+\Phi(\eta_X))/\partial \eta_V \big)d \eta_X d \eta_V$. The boundary $\partial \sigma$ is the image of the line $\eta_X$ on the plane $(\eta_X,\eta_V)$ with a shift in time $-\xi$ along the phase trajectories of the dynamic system of the system $\dot{X}=V$, $\dot{V}=-\Phi_X$ (in our case, a small value).
We assume that the boundary $\partial \sigma$ is analytically defined by the relation $\eta_V=\beta (\eta_X\:|\:\xi,\tilde{z})$ ($\eta_V <\beta$ $\forall (\eta_X,\eta_V)\in \sigma$).
Then the second term under study will take the form $\int g(\eta_X,t-\xi)F_0 ({\mathfrak E}[{\eta_X,\eta_V}])d\eta_X$.
Therefore, the right-hand side of (\ref{29}) has the form
\begin{equation}
{\mathfrak g}(\tilde{z},t;F_0)\equiv
\int_0^{\tilde{z}} g_0 (z,t)dz + \int \int^t_0 g (t-\xi,\eta_X) F_0 \big( \beta^2 (\eta_X | \xi,\tilde{z}) +
,\end{equation}
$$
\Phi (\eta_X) \big)d\xi d\eta_X
\label{30}
$$
where the tilde sign over the variable $z$ is omitted.
If we substitute into the Vlasov equation with this right-hand side (and formal annulment or replacement of the quantity ${\mathcal T}(\Phi,f)$ by an approximating term)
the normal mode of the van Kampen type ${\mathfrak K}_1(v)$, then the left-hand side will take the form $(\omega -kv){\mathfrak K}_1(v){\mathfrak K}_2(z,t)$,
and the right-hand side $i {\mathfrak y}(F_0)_v' {\mathfrak g}({z},t;F_0) \equiv {\mathfrak S}(z,t;v)$. It should be noted that this operation was enabled by the special structure of the Vlasov equation,
since the gravitational field strength here is a function closed on itself as solution to the integral equation.
Dividing both parts of the resulting equation by $(v-\omega/k)$ leads to the need to take into account an additional term, considered as a distribution exiting to the space of generalized functions
$$
f(k,v;\omega)= -k^{-1}{\mathfrak S}(z,t;v)\cdot P.V.\big(1/(\nu - v)\big)\big|_{\nu =\omega /k}+
$$
$$
\vartheta (k,\nu)\delta (\nu - v)\big|_{\nu =\omega /k},
$$
where $\vartheta(k,\nu)$ is the normalization function ($\vartheta=1+\int(-k^{-1}{\mathfrak S}(z,t;v)/(v-\nu))dv$).

The solution of the initial value problem $f(k,v,t)$ can be represented as an expansion in special solutions $f(k,v;\nu)\exp(-ik\nu t)$: $f(k,v,t)=\int {\mathcal U}(k,\nu)f(k,v;\nu)\exp(-ik\nu t)d\nu$;
accordingly, the Cauchy condition $f^{(0)}(k,v)\equiv f(k,v,t=0)=\int {\mathcal U}(k,\nu)f(k,v;\nu)d\nu$.
To determine the coefficients of ${\mathcal U}$, we obtain a singular integral equation:
$${\mathcal U}(k,v) = -k^{-1}{\mathfrak S}(z,t;v)\cdot P.V.\int\big({\mathcal U}(k,\nu)/(\nu - v)\big)d\nu + 
$$
$$
\vartheta (k,v){\mathcal U}(k,v).
$$
Its solution looks like
$$
{\mathcal U}(k,\nu)
=\frac{ {\mathcal G}_{+} (k,\nu)}{1+2\pi i {\mathcal H}_{+}(k,\nu)} -
\frac{{\mathcal G}_{-}(k,\nu)}{1+2\pi i {\mathcal H}_{-}(k,\nu)},
$$
$$
{\mathcal G}_{+}(k,\nu) - {\mathcal G}_{-}(k,\nu)={\mathcal U}(k,v),
$$
$$
{\mathcal G}_{+}(k,\nu) + {\mathcal G}_{-}(k,\nu)=\frac{1}{\pi}\int \frac{{\mathcal U}(k,\nu)}{\nu - v}d\nu,
$$
$$
{\mathcal H}_{+}(k,\nu) - {\mathcal H}_{-}(k,\nu)=-k^{-1}{\mathfrak S}(z,t;v),
$$
$$
{\mathcal H}_{+}(k,\nu) + {\mathcal H}_{-}(k,\nu) =
\frac{1}{\pi i}\int \frac{-k^{-1}{\mathfrak S}(z,t;v)}{\nu -v}d\nu.
$$
Thus, we have obtained a method for applying van Kampen waves to a formally weakly inhomogeneous system of particles  when the gravitational field strength of the complete system changes slowly.
Some explanations are required here, which are related to the presence of a cosmological term in the Liouville--Gel'fand equation.
The function ${\mathfrak S}(z,t;v)$ is defined through relation (\ref{30}) and contains the factor ${\mathfrak y}$. Recall that in the second term on the right-hand side (\ref{30}),
there is a ``full potential'' $\Phi ({\bf x})$,
which is a solution to the nonlinear Poisson Equation (\ref{7}), in which the influence of $\Lambda$-repulsion is taken into account  due to the cosmological term:
$\Delta \Phi({\bf x}) = \lambda^\dag \exp\big(-\Phi ({\bf x})/T \big)-{c^2\Lambda}$. Further, the quantity ${\mathfrak y}$ is defined as
${\mathfrak y} \equiv \lambda^{\sharp}\exp\big(-\Phi({\bf x})/T \big)$, where, in turn,
the pre-exponential factor ${\lambda}^\sharp =- \frac{\lambda^\dag}{T}\exp\big( c^2 \Lambda {\bf x}^2/T \big)$, i.e., it also depends significantly on the cosmological term.
Thus, the influence of the $\Lambda$--term on the dynamics of particles in the system under consideration is critical, and is the important factor that requires modification of standard approaches of van Kampen waves and Landau damping, as a consequence of the expansion of Landau modes in van Kampen waves.

\section{Conclusions}

The need for a more refined understanding of the possible genuine differences in the features of the early and late Universe is especially sharpened by the Hubble tension, the DESI BAO data, and other challenges. Regarding the global scale, the evolution of primordial density perturbations within various dark sector models is considered to describe the cosmic web, the voids, and large-scale filaments. On the local scale, the role of self-consistent gravitational interaction has become crucial and needs proper  
techniques to deal with and to reveal the intrinsic features of the filaments on that scale.

In this paper we considered the linearized Vlasov--Poisson equation approach, and applied the profound technique developed to analyse the wave processes in plasma physics. Namely, we showed that the van Kampen waves, associated to Landau damping and phase mixing, mark the appearance of aperiodic solutions to the linearized Vlasov-Poisson equation.  Aperiodicity then arises as an intrinsic property of the resulting filamentary structures. Of principal importance is the fact that the aperiodic structures arise when we take into consideration the repulsion of the cosmological term. The cosmological term in the local Universe description appears in view of the theorem on the identity of the gravity of the sphere and point mass within the McCrea-Milne model and weak-field General Relativity. As mentioned, that approach already made it possible to explain naturally the Hubble tension, to describe the dynamics of clusters of galaxies, and to explain the local flows. 

The appearance of aperiodicity as an intrinsic feature of the local filaments could become the subject of dedicated analyses of the observational surveys, and thus act as a probe for the role of the cosmological constant in the local Universe. Aperiodicity then has to damp upon the increase in the scale of the filaments.

\begin{acknowledgements}
We are thankful to the referee for valuable comments.
\end{acknowledgements}

\end{document}